\documentclass[11pt]{article}
\usepackage[pdftex]{graphicx}
\usepackage{wrapfig,amssymb,amsmath,multicol,subfigure,hyperref}
\usepackage[numbers, sort&compress]{natbib}
\usepackage{bm}

\newcommand{\captionfonts}{\small \it}

\makeatletter  

\long\def\@makecaption#1#2{%
  \vskip\abovecaptionskip
  \sbox\@tempboxa{{\captionfonts #1: #2}}%
  \ifdim \wd\@tempboxa >\hsize
    {\captionfonts #1: #2\par}
  \else
    \hbox to\hsize{\hfil\box\@tempboxa\hfil}%
  \fi
  \vskip\belowcaptionskip}
\makeatother   

\pdfpagewidth=\paperwidth
\pdfpageheight=\paperheight

\DeclareGraphicsExtensions{.pdf, .jpg, .tif , .ai}

\begin{document}

\pagenumbering{roman}

\centerline{\bf A Science White Paper in response to the 2012 Heliophysics Decadal Survey}
\centerline{\bf ``Solar \& Heliospheric Physics'' and ``Solar
  Wind-Magnetospheric Interactions''}
\vspace{0.2in}

\centerline{\Large \bf An  Experimental  Plasma Dynamo Program for Investigations\\[2mm]}
\centerline{\Large \bf of Fundamental Processes in Heliophysics}
\vspace{.2in}

\begin{center}
\parbox{5in}{
\begin{center}
  \mbox{Benjamin Brown (U Wisc \& CMSO)}, 
  \mbox{Cary Forest (U Wisc \& CMSO)}, 
  \mbox{Mark Nornberg (U Wisc \& CMSO)}, 
  \mbox{Ellen Zweibel (U Wisc \& CMSO)},
  \mbox{Fausto Cattaneo (U Chicago \& CMSO)} and
  \mbox{Steven Cowley (Imperial College \& CCFE, England)}
\end{center}
}
\end{center}
\vspace{0.5in}

Plasma experiments in laboratory settings offer the opportunity to
address fundamental aspects of the solar dynamo and magnetism in the
solar atmosphere.  Experiments are currently under construction that
can investigate the self-generation of magnetic fields and related
processes in large, weakly magnetized, fast flowing, and hot
(conducting) plasmas. These and future experiments will probe
questions that are of crucial importance to heliophysics in the solar
interior, atmosphere and wind. Uniquely, laboratory plasma
experiments coupled with theoretical explorations can serve to
calibrate the simulation codes which are being used to understand the
solar dynamo, magnetic reconnection and 
flares in the solar atmosphere, the nature of CMEs, and the
interactions between planetary magnetospheres and the solar wind.
Laboratory plasma experiments are likely to contribute new
understanding complementary to the traditional observational and
modeling approach normally used by space physicists.

We argue here that ground-based
laboratory experiments have direct connections to NASA based missions
and NSF programs, and that a small investment in laboratory
heliophysics may have a high payoff.  We will use the Madison Plasma
Dynamo Experiment (MPDX)\footnote{This work builds upon excitement in recent years of using liquid
  metals to study dynamos and will extend these studies to more
  astrophysically relevant parameters.  Use of a plasma for such experiments allows
  the magnetic Reynolds number (the dimensionless parameter governing
  self-excitation of magnetic fields) to be approximately a factor of
  10 larger than in liquid metal experiments.  These experiments will
  be the first to investigate self-excited dynamos in a plasma, the
  state of matter that makes up most naturally occurring astrophysical
  dynamos. }  as an example, but advocate here for broad
involvement in community-scale plasma experiments.

\paragraph{Fundamental plasma processes in heliophysics and connection to
  NASA missions and NSF programs}

  The 22-year solar cycle stands out as one of the most remarkable and
enigmatic examples of magnetic self-organization in nature.  The Sun's
cycles of magnetic activity profoundly affect our modern technological
society.  Unsurprisingly, solar magnetism is a fundamental focus of
current and future NASA missions.

  Magnetic fields that emerge at the solar surface as
  sunspots are built by dynamo action in the solar convection zone.
  In most solar dynamo models, organized fields are built in the
  tachocline, an interface layer deep in the Sun at the bottom of the
  solar convection zone.  These magnetic structures then become
  buoyantly unstable and rise through the turbulent convection to
  emerge at the surface.  After emergence, magnetic reconnection in
  the solar atmosphere can lead to flares and coronal mass ejections
  (CMEs) which have substantial impacts on the heliosphere and on
  Earth's magnetosphere in particular.  These processes of magnetic
  flux creation, emergence, and reconnection form many of the driving
  questions behind the current Solar Dynamics Observatory (SDO),
  Hinode,   the Stereo mission, the very successful Solar and
  Heliospheric Observatory (SOHO), and the upcoming ground-based NSF
  Advanced  Technology Solar Telescope (ATST).  
  Understanding the origins of the heliospheric
  magnetic field and the evolution of CMEs in the heliosphere is
  fundamental to the future Solar Sentinels mission and plays
  important roles in the future Solar Probe and Interface Region
  Imaging Spectrograph (IRIS) mission as well.
  Indeed, the evolution of solar magnetism forms the driving
  questions behind the Living With a Star (LWS) program at NASA:
  namely, \emph{how and why does the Sun vary, how does the Earth
    respond, and what are the impacts to humanity?}

  As a community, we have made significant progress in understanding
solar magnetism through a combination of theoretical treatments and
observational techniques, but many key components of the solar dynamo
remain poorly understood.  In particular, the processes behind
magnetic buoyancy and the generation of magnetic field through
turbulent correlations occur deep in the solar interior and are
difficult to constrain with either helioseismic observations or
numerical simulations.  Magnetic reconnection in the solar atmosphere
is likewise poorly understood at present.  Our understanding of
interactions between the Sun and the Earth have grown tremendously and
are greatly facilitated by in situ observations of the solar wind in
the near-Earth environment by missions like the Advanced Composition
Explorer (ACE) and the Magnetospheric Multi-scale Mission (MMM), but
many details of interactions between coherent magnetized plasma
structures like CMEs with the Earth's magnetosphere remain unclear.

\begin{figure}[t]
\includegraphics[width=\linewidth]{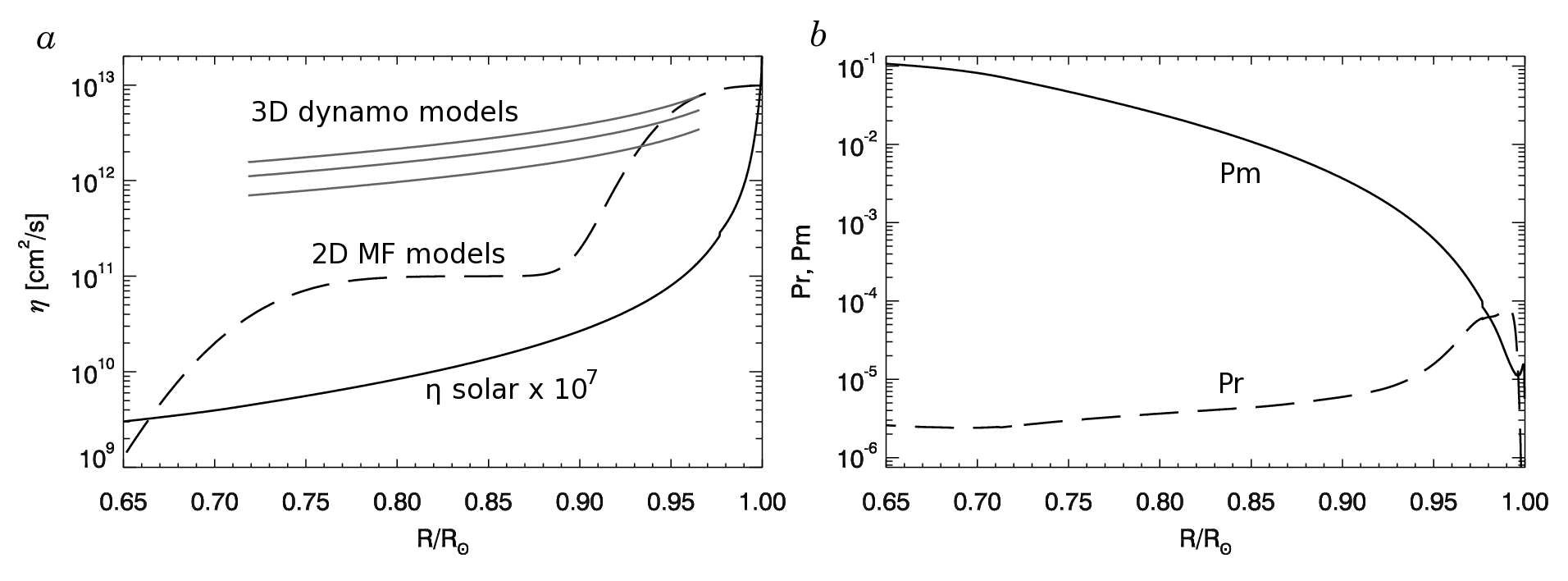}
\caption{\label{fig:eta profiles} $(a)$ Radial profiles
    of magnetic diffusivity $\eta$ used in various dynamo models.  
    Shown in grey lines are three representative profiles from 3D
    convective solar dynamo simulations using the anelastic spherical
    harmonic (ASH) code \citep[e.g.,][]{Brun_et_al_2004,
      Brown_et_al_2010b, Miesch_et_al_2010}.
    Also shown is the double-step profiles used in 2D mean-field
    Babcock-Leighton models that have been used to predict the current
    solar cycle \citep{Dikpati&Gilman_2006}. 
    The molecular diffusivity for a hydrogen plasma at
    solar conditions is also shown, multiplied by $10^7$ for display
    purposes.  $(b)$ The Prandtl number $\mathrm{Pr}=\nu/\kappa$ (dashed) and
    magnetic Prandtl number $\mathrm{Pm}=\nu/\eta$ (solid) for a
    hydrogen plasma at solar conditions.  Simulations use values of order unity.
  }
\end{figure}

A fundamental limitation facing theoretical explorations of
heliophysically relevant plasma processes is the vast separation in
parameter space between the natural phenomena and the capabilities of
even the largest super computers.  As one example, cutting edge
global-scale solar dynamo simulations run as part of the NASA High-End
Computing (HEC) program\footnote{using for example the Pleiades and Columbia supercomputers} and as
part of the NSF Partnerships in Advanced Computing
Infrastructure Teragrid program\footnote{using supercomputers at Pittsburg Supercomputing
  Center (PSC), San Diego Supercomputer Center (SDSC), Texas Advanced
  Computing Center (TACC), National Institute for Computational
  Science (NICS), at the National Center for Supercomputing
  Applications (NCSA), and others}
typically employ turbulent diffusivities which are more than ten
orders of magnitude larger than the molecular diffusivities of the
solar plasma (Figure~\ref{fig:eta profiles}$a$).  This also applies to the mean-field
models which have been used in attempts to predict amplitude and
timing of the current solar cycle \citep[e.g.,][]{Dikpati&Gilman_2006,
Charbonneau_2010}.  Yet these simulations still require substantial
computational resources and represent a significant investment by NASA
HEC and the NSF Teragrid.   

It is deeply impractical to consider simulating global-scale
dynamo action in the Sun using molecular diffusivities and doing so
would likely require a century of further growth in computational
resources.  Instead, it is vital to formulate better estimates and models
of the turbulent transport processes that occur in a plasma under
solar conditions.  As an example, even the ratio between how quickly
magnetic fields are transported and mixed by turbulence compared to
the transport of momentum by the same is poorly constrained.  Molecular
values of this magnetic Prandtl number Pm are of order
$10^{-1}$--$10^{-5}$ (Fig.~\ref{fig:eta profiles}$b$); turbulent values
are thought to be of order unity, but this has not been adequately
explored for real turbulent transport in a stratified plasma.  These
problems are not unique to the solar interior, but are true for
simulations of photospheric convection that couple to the chromosphere
and corona \citep[e.g.,][]{Nordlund_et_al_2009}, and simulations that
capture the interactions between the solar wind, CMEs and planet
magnetospheres throughout the heliosphere and to the boundary with
interstellar space itself.

The growth of computational resources has made it feasible to directly
simulate the plasma regime explored by experiments like MPDX.  A
program of laboratory experimentation that is tightly coupled with
significant numerical simulation efforts will lead to new insights
into key plasma processes.  In particular, these laboratory plasma
experiments will likely achieve self-sustaining dynamo action, will
explore the dynamics of magnetic buoyancy in a stratified plasma
atmosphere, and will explore the interactions between an Earth-like
magnetosphere and magnetotail with a solar-wind like plasma flow.  In
addition to revealing the fundamental underlying plasma processes at
work in these systems, these experiments can anchor the
computationally intensive simulations which are used to model the
solar dynamo, the buoyant rise and emergence of magnetic structures in
the solar interior, and interactions of the solar wind and CMEs with
the Earth's magnetosphere.

\paragraph{The importance of simulations coupled to experiments}
Just as we cannot directly simulate the turbulent conditions of solar
convection, we cannot construct laboratory experiments that capture
the vast spectrum of scales present in the Sun.  Some important
physical ingredients acting in solar convection are very difficult to
reproduce in experiments: in particular, it is very difficult to
produce conditions that mimic the radial self-gravity of systems like
the Sun, and experiments that do must generally be conducted at great
cost in low-g environments \citep[e.g.,][]{Hart_et_al_1986}. 

Laboratory experiments provide great opportunities for  exploring key
physical processes occurring in plasmas under solar conditions, but
they cannot capture the integrated global-scale picture.  Global-scale
couplings between convection, magnetism and rotation are likely
crucial to fully understanding the solar cycle as a whole and may be
important for understanding eruptive flares and CMEs as well.  
At present, simulations are the only option for exploring the coupled
global-scale system.  Simulations can offer insight into processes
that are difficult or impossible to directly measure in the laboratory
(e.g., the full 3-D velocity field and magnetic field at all points in
a volume at all instances in time) and can suggest profitable avenues
for additional experimental exploration.  But this is not a one-way
process: the laboratory experiments validate the codes and inform those
simulations in how to correctly capture processes that currently elude
simple theoretical explanation, including turbulent transport,
magnetic buoyancy instabilities, and the non-linear couplings that
arise between flows and magnetism.

Ultimately, theoretical explorations, numerical simulations and
laboratory experiments must work together in a tightly coupled fashion
to achieve optimal results.  But this is feasible and can yield large
and rapid payoffs.
In the MPDX experiment for example, this work has begun already for
modeling of  the dynamo and magneto-rotational instability scenarios
using the NIMROD code, which solves the MHD equations with 
a number of two-fluid extensions.  These simulations have indicated
that sustained dynamo action is possible and in which parameter
regimes this is most achievable.  New simulations are exploring the
possibility of buoyancy instabilities, including magnetic buoyancy,
and suggest that these may be experimentally achievable.
Direct comparisons between the numerical simulations
and pathfinder prototype experiments  are completing the code 
{\sl validation} loop even before construction finishes on the main
experiment (MPDX).  Theoretical explorations of the simulations and
the experiment are in turn providing the ultimate link to the plasma astrophysics.

Tightly linked programs of laboratory experimentation and numerical
simulation, backed by significant theoretical analysis to understand
both systems, offer opportunities for rapid advancement in our
understandings of solar plasma physics.

\begin{figure}[t]
\begin{center}
\includegraphics[width=6.5in]{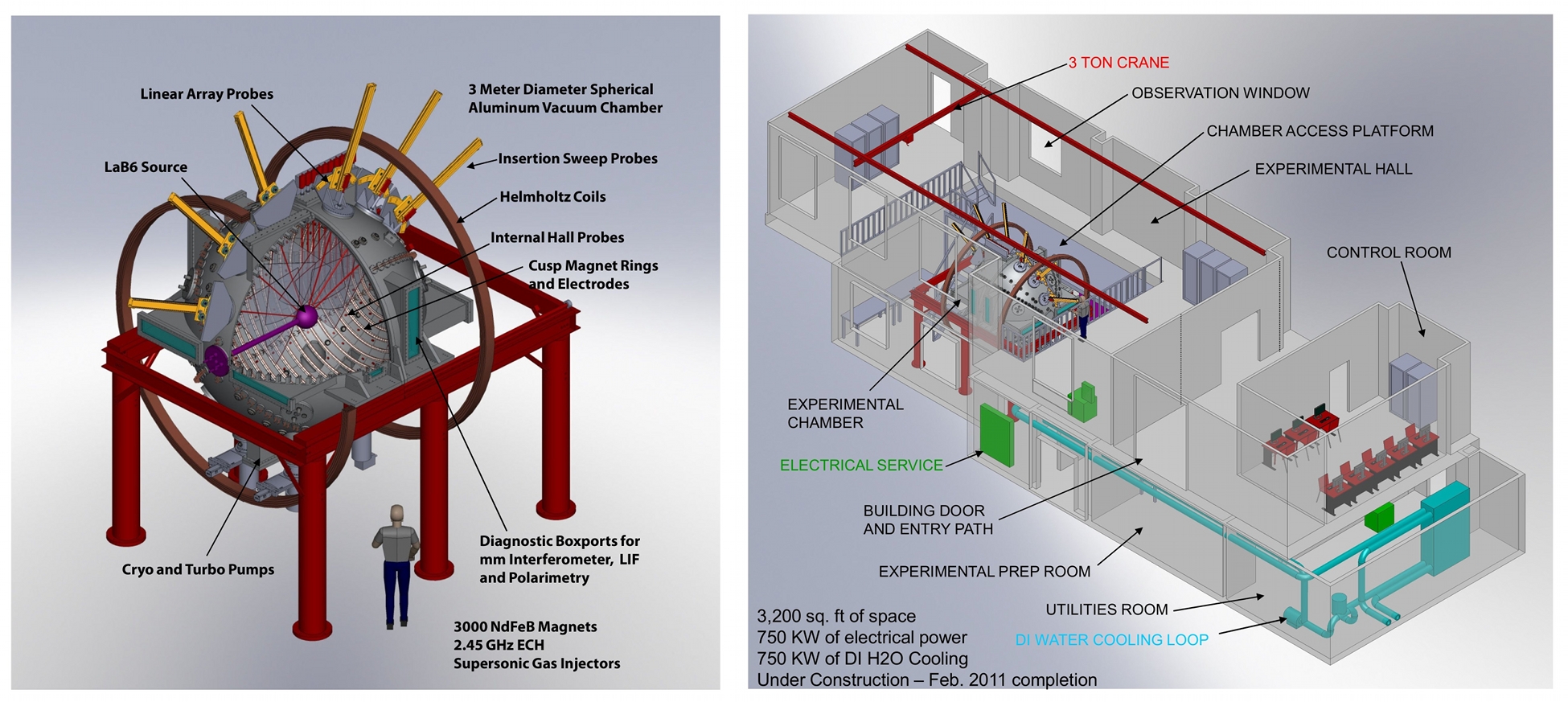}
\end{center}
\caption{The Madison Plasma Dynamo Experiment (MPDX), with human
 figure to scale (left).  
 Plasma Dynamo Laboratory Overview (right).  The drawing shows the
 new Plasma Dynamo Facility under construction at the UW Madison.
 The main experimental hall is two
 stories high, with approximately 2000 ft$^2$ of floor space and houses the 3 m
 diameter Plasma Dynamo Experiment. 
 \label{fig:MPDX}
}
\end{figure}

 \paragraph{Status of MPDX and future plasma experiments}  

Compared to space missions, ground based laboratory experiments can be
quite affordable and can be built on rapid timescales.  As~an~example,
the MPDX experiment has been funded for construction through the NSF
Major Research Instrumentation program with at total device cost of
\$2.5M (part of which was \$750k in cost-sharing from the UW).  In
2009 a Plasma Dynamo Kick-Off Meeting, sponsored by the NSF Center for
Magnetic Self-Organization,\footnote{\tt[http://cmso.info/html/meetings/wi\_dec09.htm]}
was held to help formulate the scientific program and to provide
feedback on the design of the experiment.  Funding began in 2009, and
the 3 year project is on schedule: 2010 was spent on design and lab
remodeling, in 2011 the machine will be assembled, and in 2012 the
experiment should be ready for initial operations.  An illustration of
the experimental apparatus and the Plasma Dynamo Laboratory built
around it is shown in Figure~\ref{fig:MPDX}.

There are significant barriers to running such programs on a community
scale.   Using MPDX as an example again, after construction there is not yet a
program in place to operate this experiment.  This experiment is
larger than a typical single-investigator NSF experiment and might be
considered a {\sl medium-scale} basic plasma or laboratory plasma
astrophysics experiment as called for in the recent Plasma 2010 report
from the National Research Council.  The operations of such an
experiment and the diagnostic requirements are similar in scale to
university-scale experiments in fusion research, yet there is no
mechanism for funding the operation of a community-based facility such
as this.  A major challenge is adequately funding the operations of such
experiments and also the significant theoretical explorations and numerical
simulation efforts which are vital to maximizing the scientific
returns of such facilities.

\paragraph{Heliophysically relevant plasma experiments}  

Future plasma experiments could directly address many plasma processes
that are directly relevant to heliophysics and NASA.  Most
experimental plasma facilities at present are focused on other
science, with substantial investments in exploring basic plasma
processes and the application to fusion energy generation.  Plasma
experiments can now directly address several key components of the
solar dynamo and of eruptive events in the solar atmosphere.  

A variety of laboratory experiments are feasible in the near future that are clearly
heliophysically relevant: 
 \begin{enumerate}
      \item large-scale dynamo experiments (generation of the large scale solar magnetic field), 
      \item small-scale turbulent dynamo experiments (small scale structure of the solar dynamo and surface magnetism), 
      \item buoyancy driven stratified convection (solar convection zone and surface),
      \item magnetic buoyancy instabilities in a stratified atmosphere (sunspot emergence and solar tachocline physics),
      \item flow driven plasma turbulence in near-equipartition (solar wind),
      \item flow driven plasma turbulence impacting a model magnetosphere (space weather, sun-earth connection),
      \item magnetic field line stretching with potentially explosive, flow-driven  reconnection (magnetotail, solar flares),
      \item plasma instabilities at high $\beta$ and low collisionality (solar wind spectrum).
  \end{enumerate}
A program of heliophysically relevant laboratory plasma experiments
can provide important insight into these problems.  Such a program
will surely suggest additional unexpected avenues for research, as
expertise is gained in experimental techniques, and in the theoretical
and numerical explorations of such experiments.

The proposed plasma experiments transcend the scope of \emph{single user facilities}
and instead require substantial \emph{community-based facilities}.
Such experiments should be supported at a larger scale than can be
undertaken by individual universities and research groups.
This support must extend past the mere construction of experimental
facilities: instead, significant advances can be achieved by ongoing
support for operations, for theoretical explorations, and for tightly
coupled numerical simulations.  These synergistic efforts are likely
to maximize scientific output and may rapidly advance our
understanding of plasma processes occurring in the solar interior,
atmosphere and wind, whereas isolated efforts must continue
to wrestle with intractable limitations.
In summary, we advocate that heliophysics could greatly benefit from
the support of  community-scale, heliophysically relevant plasma experiments.

\newcommand\apj{{ApJ}}%
\newcommand\apjl{{ApJ}}%
\paragraph{References} ~\\[-2cm]

\providecommand{\newblock}{}

\end{document}